\documentclass[12pt,twocolumn,doublespace]{iopart}

\usepackage{iopams}  
\usepackage{epsfig}
\usepackage{graphics}
\usepackage{dcolumn}
\usepackage{bm}
\usepackage{color}

\begin{document}

\title[Unusual persistence of superconductivity against high magnetic fields]
{
Unusual persistence of superconductivity against high magnetic fields in the strongly-correlated iron-chalcogenide film FeTe:O$_{x}$ 
}

\author{I K Dimitrov$^{1}$,\footnote{contributed equally},\footnote{corresponding author}, W D Si$^{1}$,\footnote{contributed equally},\footnote{corresponding author}, W Ku$^1$, S J Han$^1$, and J Jaroszynski$^2$}

\address{$^1$ Condensed Matter Physics and Materials Science Department, Brookhaven National Laboratory, Upton, New York 11973-5000, USA}

\address{$^2$ National High Magnetic Field Laboratory, Florida State University, Tallahassee, Florida 32310, USA}

\ead{idimitrov@bnl.gov}
\ead{wds@bnl.gov}

\date{\today}

\begin{abstract}

We report an unusual persistence of superconductivity against high magnetic fields in the iron chalcogenide film FeTe:O$_{x}$ below $\approx 2.5$ K.  Instead of saturating like a mean-field behavior with a single order parameter, the measured low-temperature upper critical field increases progressively, suggesting a large supply of superconducting states accessible via magnetic field or low-energy thermal fluctuations.  We demonstrate that superconducting states of finite momenta can be realized within the conventional theory, despite its questionable applicability.  Our findings reveal a fundamental characteristic of superconductivity and electronic structure in the strongly-correlated iron-based superconductors.

\end{abstract}

\maketitle

\section{Introduction}

Ever since the discovery of iron-based superconductors in 2008 \cite{Kamihara}, an avalanche of research effort was aimed at understanding the pairing mechanisms behind the origins of superconductivity in these materials \cite{Mazin}.  These novel systems have shown similarities and differences to both high ${T}_{c}$ and heavy fermion systems \cite{Mazin}.  What has made the study of iron-based materials even more elusive is that some experimental data suggest that the antiferromagnetic and superconducting order parameters compete in certain systems, while they coexist in others \cite{Zhao, Lester, Nandi, Chauviere, Hardy}.  In very recent times it has been suggested that magnetic and structural instabilities can potentially form the superconducting state \cite{Mazin}.   

However, in order to understand the nature of the superconducting state and its underlying mechanism, it is imperative to gain insight into a materials electronic structure \cite{Tinkham,DeGennes}.  One of the most relevant approaches to studying the electronic structure of superconductors has come in the way of exploring their magnetic responses to an external magnetic field, \emph{H}, as a function of temperature, \emph{T}.  In the case of BCS superconductors, the upper critical field, ${H}_{c2}$, has been used as a measure of electronic coherence \cite{Tinkham}, as well as an indicator of the relevant pairbreaking mechanisms \cite{Tinkham,Werthamer,Clogston}.  Detailed analyses of the $H-T$ phase diagrams of iron pnictides and chalcogenides have been utilized to suggest that they are multi-band superconductors with unconventional pairing mechanisms \cite{Khim,Cho}.

Here, we report a striking persistence of superconductivity against high magnetic fields in the iron chalcogenide film FeTe:O$_{x}$ below 2.52 K.  The measured low-temperature ${H}_{c2}$ increases progressively, showing a pronounced inflection, instead of the saturation expected from a mean-field theory with a single order parameter.  This suggests the presence of a large supply of superconducting states accessible via magnetic field.  Additionally, our observations suggest that the rapid reduction of the upper critical field with increasing temperature (concomitantly with the superfluid density) is a consequence of thermal fluctuations involving these states.  We explore a scenario of Cooper pairs with finite center-of-mass momentum and find that, while not perfectly justifiable, it is consistent with our data below $T \approx 2.5$ K.  Our findings reveal a fundamental characteristic of superconductivity and electronic structure in the strongly-correlated iron-based superconductors.

The procedure for the FeTe:O$_{x}$ film deposition was described in detail elsewhere \cite{Weidong}, albeit the lower ${T}_{c}$ of the present sample compared to the one discussed by Si \emph{et al.} \cite{Weidong} (a zero-field temperature scan of the resistivity exhibits a ${T}_{c}$ of 7.05 K, when in order to obtain the superconducting transition we consider the mid-point of the normalized resistivity, 50$\%$ of $\rho \left( T \right ) / {\rho}_{0} \left ( T \right)$).  Currently, the mechanism of oxygen-induced superconductivity in FeTe:O$_{x}$ films is still unknown \cite{Weidong,QL}.  Further investigations to sort out the effects of oxygen in this class of films present an important scientific objective for the field.  Here, however, we focus exclusively on the behavior of superconductivity due to the presence of an external magnetic field.

\section{Experimental Techniques and Data Analysis Details}

ac transport measurements of the parallel and perpendicular components of ${H}_{c2}$ with respect to the crystallographic \emph{c} axis, ${H}_{c2}^{||c}(T)$ and ${H}_{c2}^{\perp c}(T)$, were carried out at the National High Magnetic Field Laboratory in Tallahassee, Florida in dc magnetic fields from 0 to 35 T.  The field was applied at a rate of 5 T/min during increasing field ramps at fixed temperatures.  The resistivity, $\rho(T)$, was measured via phase-sensitive lock-in detection (\emph{I} = 5 $\mu$A; \emph{f} = 17 Hz) at a variety of $T$'s from 1.5 -- 8 K.  The data from the field scans were normalized with respect to the normal state resistivity, $\,{\rho}_{0}(T)$, obtained from temperature scans at $H$ = 35 T, from 15 K to 7 K.  The normalized resistivities for $H \parallel c$ and $H \perp c$ at select temperatures are shown in figure 1(a) and (b), respectively \cite{Ivo1}.

\begin{figure}[h]
 	\begin{center}
		\includegraphics[width=0.47\textwidth]{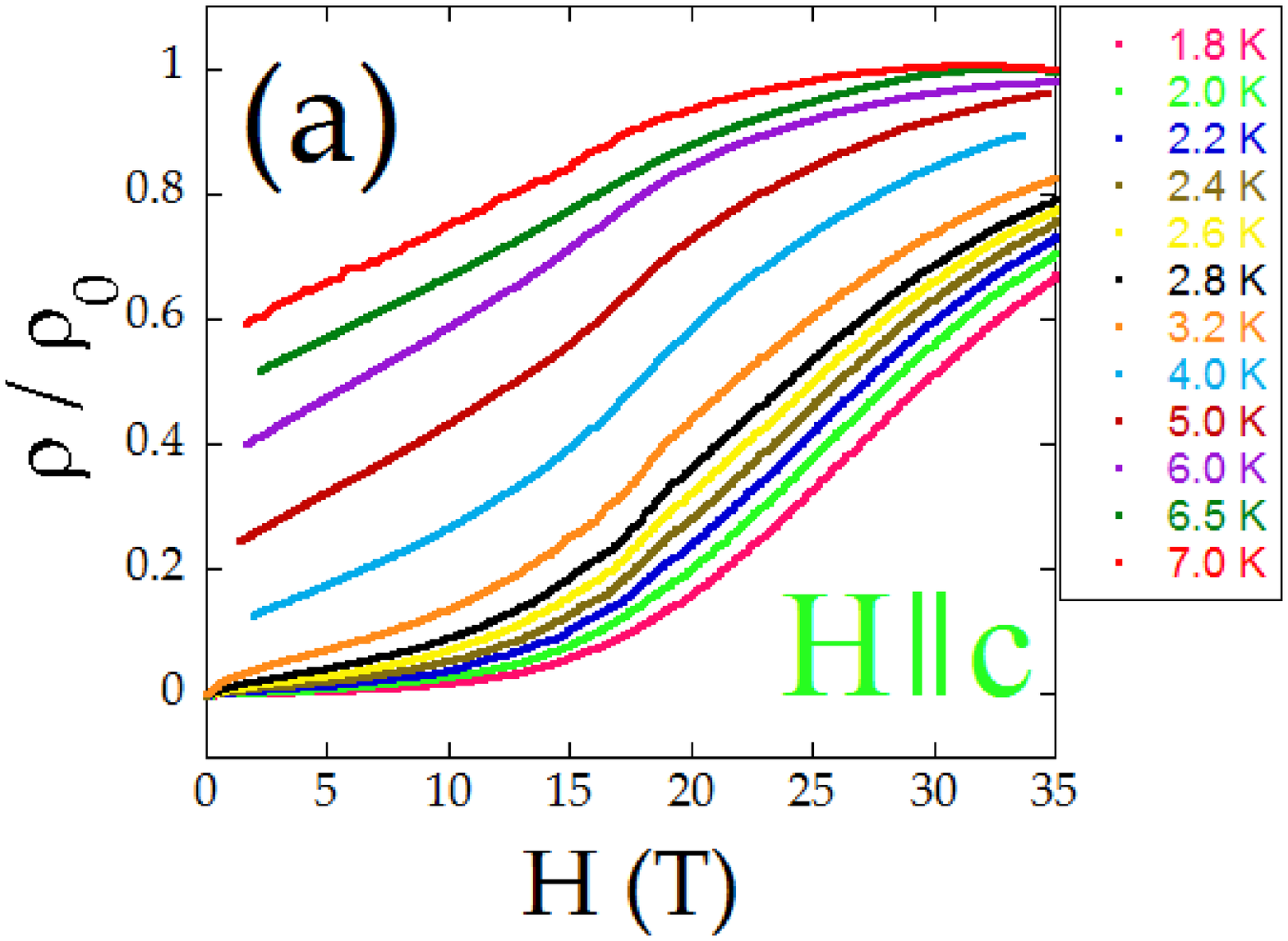}
		\includegraphics[width=0.47\textwidth]{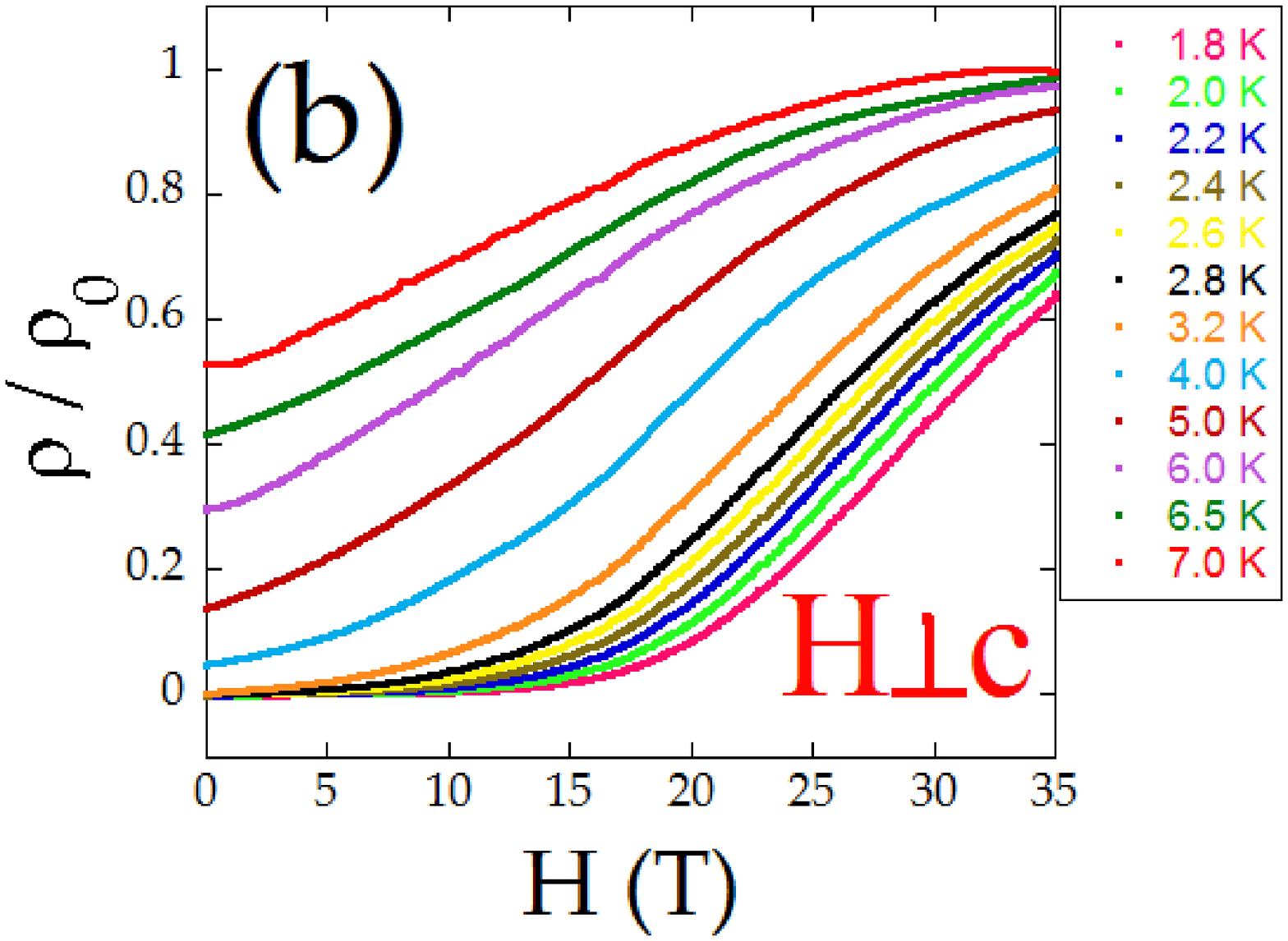}
		\includegraphics[width=0.44\textwidth]{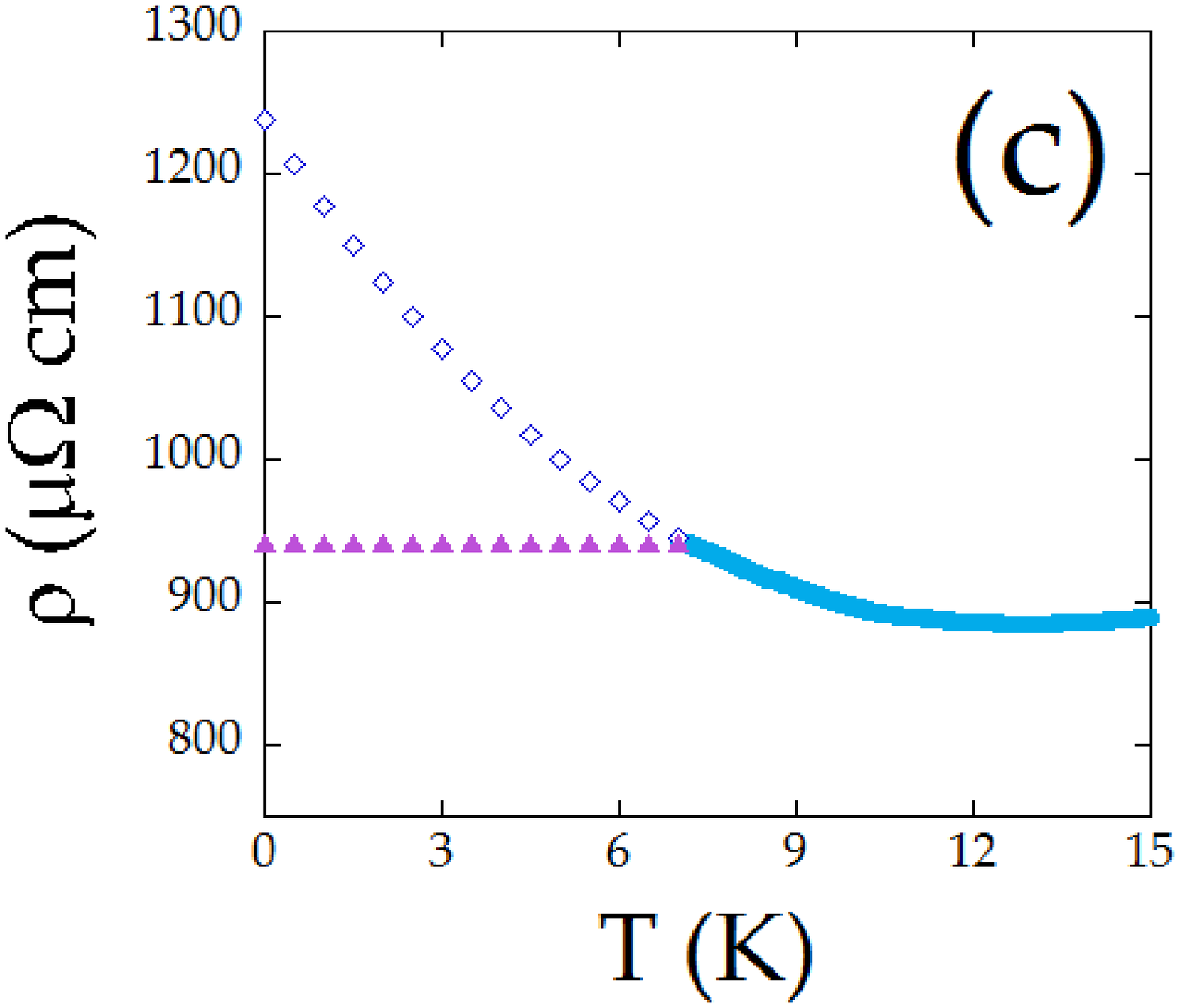}
		\caption{Magnetic field dependence of the normalized resistivity of FeTe:O$_{x}$ at select temperatures with (a) $H \parallel c$ and (b) $H \perp c$.  ${\rho}_{0}$ denotes the sample resistance of the normal state at \emph{H} = 35 T, while $\rho$ is the measured resistance.  ${\rho}_{0}$ is defined according to the temperature scan from 15 K to 7 K at 35 T, shown in light blue in (c).  The definition for ${\rho}_{0}(T)$ for the data analysis below 7 K is obtained from a third-order polynomial fit through the 7 -- 15 K data (shown in dark-blue empty rhombi in (c)).  The data is also analyzed using an alternative definition for ${\rho}_{0}$, which assumes a constant ${\rho}_{0}$ below 7 K (depicted in the straight line of filled purple triangles in (c)) \cite{Ivo1}.}
	\end{center}
\end{figure}

The mid-point of the normalized resistivity, ($50\%$ of $\rho(T)/{\rho}_{0}(T)$), was chosen to define the experimental ${H}_{c2}(T)$ \cite{Ivo2}, since the 50$\%$ criterion allowed us to analyze the full range of the data set.  $H \perp c$ and $H \parallel c$ are shown in figure 2.  We point to a very steep increase in the ${H}_{c2}$ curves near ${T}_{c}$ ($-d{H}_{c2}^{\perp c}/dT|_{H \leq 10 T} \approx$ 9.24 T/K and $-d{H}_{c2}^{\parallel c}/dT|_{H \leq 10 T} \approx$ 6.06 T/K).  It is also to be noted that $\rho(T,35$T$)$ in our sample exhibits an insulator-like behavior right above the superconducting transition (figure 1(c)), despite the fact that it has a Fermi surface and is expected to behave like a metal.

\section{Theoretical Analysis and Discussion}

The most remarkable observation here is the unusual persistence of superconductivity against high magnetic fields at low temperatures, manifest in an inflection of the phase boundary of the $H-T$ phase diagram (figure 2), which is clearly present in the $H \perp c$ case, and less drastically in the $H \parallel c$ one.  Apparently, at low temperatures ($T \leq 2.5$ K), the superconductivity survives much stronger fields by readjusting itself in some non-trivial way.  This is in great contrast to the phase diagram saturation characteristic of a BCS superconductor described by a mean-field theory of a single order parameter.  Generally speaking (without involving any specific theory), this indicates that there is a large number of low-energy superconducting states of different momenta (in a flat energy landscape) accessible via magnetic field.  The thermal fluctuations involving these states cause the rapid reduction of ${H}_{c2}$ (and the superfluid density) as the temperature increases.  Given the large gap size of the system ($\Delta\sim 3.5 k_BT_c$) \cite{Nakayama,Homes}, such fluctuations are likely in the phase (not amplitude) of the order parameter.  This strong coupling behavior is expected in the underdoped samples \cite{Luetkens} (like the present one) with correlated electronic structure \cite{Ku,Kotliar}, as evident from the proximity to magnetic/orbital/structural orders.

\begin{figure}
	\begin{center}
		\includegraphics[width=0.85\textwidth]{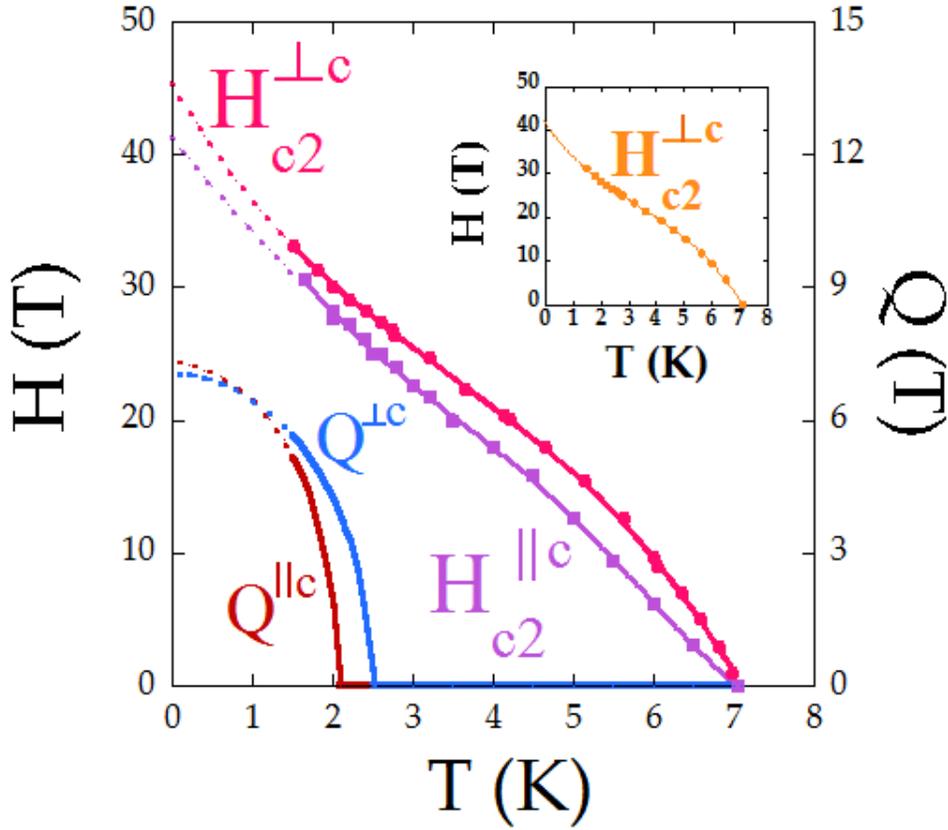}
			\caption{Phase diagram constructed from the normalized resistance versus temperature data.  The ${H}_{c2}$(\emph{T}) for $H \perp c$ (pink solid circles) and $H \parallel c$ (violet solid squares) were defined at 50$\%$ of ${\rho}_{0}(T)$ from the field scans (figure 1(a) and (b)).  The dotted lines below \emph{T} = 1.5 K are guides for the eye.  The calculated values of ${Q}^{\perp c}(T)$ and ${Q}^{\parallel c}(T)$ are shown in red and blue, respectively.  Inset: Shown is the $H-T$ phase diagram for $H \perp c$ obtained by the alternative analysis condition, setting ${\rho}_{0}$ to be a constant below 7 K (figure 1(c)) \cite{Ivo1}.}
	\end{center}
\end{figure}	

Nonetheless, due to the poor understanding of such a strong coupling regime, we stay within the conventional framework of BCS theory, as this is the only currently-available approach.  The Werthamer-Helfand-Hohenberg (WHH) theory was developed to describe the orbital-limited upper critical field, ${H}_{c2}^{O}(T)$, for a single active band in the weakly-coupled limit: ${H}_{c2}^{O}$(0) = --0.69$dH/dT|_{T = {T}_{c}} {T}_{c}$ \cite{Werthamer}.  Applying it to our FeTe:O$_{x}$ thin film $H-T$ phase diagram yields ${H}_{c2}^{O}(0)$ = 44.95 T and 29.48 T in $H \perp c$ and $H \parallel c$, respectively.  On the other hand, the expected paramagnetic pair-breaking (Pauli-limited) upper critical field, ${H}_{c2}^{P}(0)$, in the same sample is estimated to be 1.86 ${T}_{c}$ = 13.11 T from the Clogston-Chandrasekhar (CC) theory \cite{Clogston}.  In a related transport measurement, Khim \textit{et al.} argued that the much larger ${H}_{c2}^{O}$'s in comparison to the ${H}_{c2}^{P}$ measured in a FeTe$_{0.6}$Se$_{0.4}$ single crystal suggests that these observations could imply the importance of paramagnetic pair-breaking effects, or be due to multi-band scattering in the latter system \cite{Khim}.  Notwithstanding the above interpretation, both WHH and CC are derived in the weakly-coupled limit, while their validity in a strongly coupled, non-BCS system such as Fe(Te,Se) remains dubious \cite{Kato,Hanaguri,Khim}.    

Within BCS, the finite momentum superconductivity is explained within the FFLO picture \cite{Larkin,Fulde}.  This should be considered as one possible scenario consistent with the above general considerations, but not necessarily representing the only (or even the correct) microscopic picture.  The FFLO state is marked by a smaller condensation energy, but also by a lower Zeeman energy, compared to the normal state, resulting in an overall suppression of the normal state at magnetic fields higher than ${H}_{c2}^{P}$ = 2$\Delta$/\emph{g}${\mu}_{B}$ (Clogston limit \cite{Clogston}), where $\Delta$, \emph{g} and ${\mu}_{B}$ stand for the superconducting gap, the gyromagnetic ratio of a free electron, and the Bohr magneton, respectively \cite{Larkin,Fulde,Matsuda}.  The FFLO state has been reported in a number of organic \cite{Matsuda,Tanatar}, and heavy fermion superconductors \cite{Matsuda,Gloos}.  In general, strong type II superconductors with large Maki parameters, ${\alpha}_{M} = \sqrt{2}{H}_{c2}^{O}/{H}_{c2}^{P}$, in the clean limit, $\xi \ll l$, are the typical candidates for FFLO state systems, where $\xi$ and $l$ denote the superconducting coherence length and the electronic mean free path, respectively \cite{Matsuda}. 

Subsequently, we set out to analyze the \emph{H--T} phase diagram of FeTe:O$_{x}$ using the established FFLO theoretical framework \cite{Gurevich}.  We minimized the equation of state for the parameter $q \left( T \right)|_{H = {H}_{c2}}$ of a two-band model which accomodates orbital and paramagnetic pair-breaking mechanisms (equation (1)) at each point in the \emph{H} versus \emph{T} phase diagram \cite{Gurevich}.  Finally, we used this parameter to derive $Q(T)|_{H = {H}_{c2}}$ \cite{Gurevich}:
\begin{eqnarray}
a_{1}(ln\,t + U_{1}) + a_{2}(ln\,t + U_{2}) + (ln\,t + U_{1})(ln\,t + U_{2}) = 0
\end{eqnarray}
where \emph{a$_{1}$} = ($\lambda_{0}$ + $\lambda_{-}$)/2$w$, \emph{a$_{2}$} = (${\lambda}_{0}$ -- ${\lambda}_{-}$)/2$w$, with ${\lambda}_{-}$ = ${\lambda}_{11}$ -- ${\lambda}_{22}$, ${\lambda}_{0}$ = (${\lambda}_{-}$ + 4${\lambda}_{12}{\lambda}_{21}$)$^{1/2}$, and $w$ = ${\lambda}_{11} {\lambda}_{22}$ -- ${\lambda}_{12}{\lambda}_{21}$.  ${\lambda}_{ij}$ (\emph{i, j} = 1, 2) define the coupling constants used in the theory \cite{Gurevich}.  In the limit of strong interband pairing (${\lambda}_{12} {\lambda}_{21} \gg {\lambda}_{11}{\lambda}_{22}$), we obtain ${\lambda}_{11}$ = ${\lambda}_{22}$ = 0 and ${\lambda}_{12}$ = ${\lambda}_{21}$ = 0.5, leading to ${a}_{1}$ = ${a}_{2}$ = --1 \cite{Gurevich}. 
In the above equation we used the integral form for \emph{U$_{1}$} \cite{Gurevich}:
\begin{eqnarray}
U_{1} = \ln(4\gamma) + \frac{te^{q^{2}}}{\sqrt{b}} \int_q^\infty \mathrm{d}{u}   \;    e^{-u^{2}} \times
\Im \, \left\lbrace \ln \frac{\Gamma \left[ 1/2 + i \left( \alpha b + u \sqrt{b} \right)/t \right]}{\Gamma \left[ 1/2 + i \left( \alpha b - u \sqrt{b} \right)/t \right]} \right\rbrace
\end{eqnarray}
where $\gamma \approx 1.78$ and \emph{t} = \emph{T/T$_{c}$}.  The rest of the quantities are defined as follows:
\begin{eqnarray}
q = {\left( \frac{{Q}_{z}^{2}{\phi}_{0}{\epsilon}_{1}}{2 \pi H} \right)}^{1/2} \; , \qquad
b = \frac{{\hbar}^{2} {v}_{1}^{2} H} {8 \pi {\phi}_{0} {k}_{B}^{2} {T}_{c}^{2} {\omega}_{1}^{2}} \; , \qquad
\alpha = \frac{4 \mu {\phi}_{0} {k}_{B} {T}_{c} {\omega}_{1}}{{\hbar}^{2}{v}_{1}^{2}}
\end{eqnarray}
where ${Q}_{z}$, ${\phi}_{0}$ and $\mu$ stand for the projection of vector \emph{Q} along the field direction, flux quantum and the magnetic moment of a quasiparticle, respectively, while ${\epsilon}_{1}$, ${v}_{1}$, ${\omega}_{1}$ represent the mass anisotropy, Fermi velocity and the Eliashberg constant, ${\omega}_{1}$ = 1 + ${\lambda}_{11}$ + $|{\lambda}_{12}|$, for band \emph{1}.  ${U}_{2}$ is obtained by replacing ${\omega}_{1}$ with ${\omega}_{2}$ = 1 + ${\lambda}_{22}$ + $|{\lambda}_{21}|$,
\emph{b} with \emph{b}$\eta$ (except terms $\propto \alpha b$), and \emph{q} with \emph{q}$\sqrt{s}$ in ${U}_{1}$.  Here, $\eta$ = ${v}_{2}^{2}/{v}_{1}^{2}$, $s$ = ${\epsilon}_{2}/{\epsilon}_{1}$, and ${v}_{2}$ and ${\epsilon}_{2}$ represent the Fermi velocity and mass anisotropy for band \emph{2}, respectively \cite{Gurevich,Cho}.  The Fermi velocity used in the calculation, ${v}_{1}$ = 0.7 eV$\cdot$\AA = 1.0635 $\times$ 10$^{5}$ m/s, was obtained from an ARPES measurement on a Fe$_{1+x}$Te single crystal \cite{Xia}.  $\mu$ = 2.2${\mu}_{0}$,\cite{Kotliar} where ${\mu}_{0}$ is the Bohr magneton, in order to calculate $b \left( T, H \right)$ and $\alpha$.  ${\epsilon}_{1}$ is related to the anisotropy parameter, ${\gamma}_{H}$, by ${\epsilon}_{1} = 1/\sqrt{{\gamma}_{H}}$, and ${\gamma}_{H}(T)$ was derived from ${\gamma}_{H}(T) = {H}_{c2}^{\perp c}(T) / {H}_{c2}^{\parallel c}(T)$ (figure 3) and taken to be equal to 1 for convenience.  The field and mass anisotropies, ${\gamma}_{H}$ and ${\gamma}_{m}$, cannot be necessarily considered equal as in the case of the anisotropic single-band superconductors \cite{Kogan,Kidszun} and the discrepancy between the two has been interpreted to be a signature of multi-band physics in a LaFeAsO$_{1-x}$F$_{x}$ oxypnictide film \cite{Kidszun}.  Furthermore, we performed an angular-dependent transport measurement of ${H}_{c2}(T$ = 1.75 K) (figure 3).  ${H}_{c2}(\phi,1.75\,K)$ is rather nicely fitted with a calculation based on the single band anisotropic Ginzburg-Landau theory (inset of figure 3) \cite{Blatter}.  Finally, we approximated the ratio of the mass anisotropies in the two bands as \emph{s} = 1 while $\eta$ = 0.3 \cite{Cho}.  $\alpha$ was calculated to be 0.016(${\Delta}^{w}$/${k}_{B}$), where ${\Delta}^{w}$ is the superconducting gap in the weakly-coupled limit (${\Delta}^{w} \approx 1.74 {k}_{B} {T}_{c}$).  However, a typical gap observed in FeTe superconductors is on the order of 3.5${k}_{B} {T}_{c}$ \cite{Nakayama,Homes}.  Thus, in order to get the correct value for $\alpha$ we used $\Delta$ = 3.5${k}_{B} {T}_{c}$ instead, finally obtaining $\alpha$ = 0.4.  No further rescaling of the parameters \emph{q}, $\alpha$ and \emph{b} was pursued.  

The results of our calculation for ${Q}^{\perp c}(T)|_{ H = {H}_{c2}}$ and ${Q}^{\parallel c}(T)|_{ H = {H}_{c2}}$ are shown in units of 1/$\sqrt{{\phi}_{0}}$ in figure 2.  Remarkably, we obtain ${H}_{c2}^{\perp c}(0) \approx$ 42 T and ${H}_{c2}^{\parallel c}(0) \approx$ 44 T upon calculating the ${H}_{c2} \left( 0 \right)$ in the paramagnetic limit, $Q(0) \approx 2 \mu H/ \hbar {v}_{F}$ \cite{Matsuda}.  ${H}_{c2}^{\perp c}(0)$ obtained from ${Q}^{\perp c}$(T)$|_{ H = {H}_{c2}}$ agrees nicely with the guide for the eye drawn in figure 2 for ${H}_{c2}^{\perp c}(0)$, while the respective result for ${H}_{c2}^{\parallel c}(0)$ suggests that the ${H}_{c2}^{\parallel c}$(T) curve is slightly steeper than the guide for the eye.  It is to be noted that ${Q}^{\perp c}$(T)$|_{ H = {H}_{c2}}$ starts to exhibit non-zero values at 2.52 K, just above the unusual upturn in ${H}_{c2}^{\perp c}(T)$, which translates to 0.36$\,{T}_{c}$.

\begin{figure}[h]
	\begin{center}
		\includegraphics[width=0.75\textwidth]{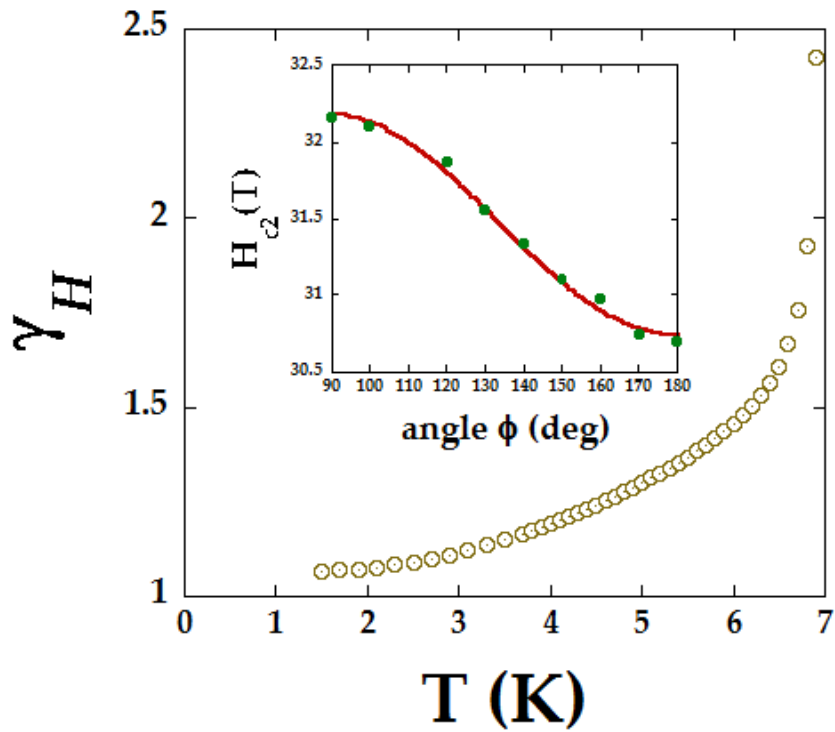}
			\caption{The anisotropy parameter, ${\gamma}_{H} = {H}_{c2}^{\perp c}/{H}_{c2}^{\parallel c}$, as a function of temperature, obtained from ${H}_{c2}^{\perp c}$ and ${H}_{c2}^{\parallel c}$, shown in Fig. 2.  Inset: shown is the angular dependence of ${H}_{c2}(T)$ at $T$ = 1.75 K from $\phi = {90}^{\circ}$ ($H \perp c$) to $\phi = {180}^{\circ}$ ($H \parallel c$) (in filled green circles), obtained from 35 -- 0 T field scans (d$H$/d$T = -7$ T/min).  The solid red line is a fit of ${H}_{c2}(\phi) = {H}_{c2}^{\perp c} \cdot$[cos$^{2} \left( \phi \right) +$ sin$^{2} \left( \phi \right)/{{\gamma}_{m}}^{2} \left( T \right)$]$^{-0.5}$ through the data points, yielding ${H}_{c2}^{\perp c}(1.75\,K)$ = 30.75 T and ${\gamma}_{m}(1.75\,K)$ = 1.05.}
	\end{center}
\end{figure}

We note, however, that the measured ${\gamma}_{H}(T)$ is very close to unity, especially in the temperature range where the calculated $Q(T)$ acquires non-zero values (figure 3).  The lower anisotropy is not favorable for the formation of an FFLO state \cite{Gurevich}, and especially surprising is the fact that we measure non-zero $Q(T)$ which coincides with a very unusual behavior in ${H}_{c2}(T)$ in a system which is not in the clean limit.  Also, the weakly-coupled two band scenario entertained by Kidszun \textit{et al.} \cite{Kidszun} would be inconsistent with the strong intraband scattering and Pauli-limited ${H}_{c2}$'s present in iron-based superconductors \cite{Gurevich}.

All of these suggest that while the general physical considerations concerning the large supply of superconducting states is reasonable, the detailed microscopic understanding remains elusive, and likely to be enriched by the strongly-correlated nature of the electronic structure of this system.  The observed unusual persistence of superconductivity against high magnetic fields not only reveals a fundamental characteristic of the superconductivity in this system, but also puts a strong constraint on the microscopic understanding of the electronic structure of this new class of superconductors.

\section{Summary and Outlook}
In conclusion, we have found a striking persistence of superconductivity in a FeTe:O$_{x}$ thin film at high magnetic fields, especially when the field is applied perpendicular to the crystallographic \emph{c} plane.  The upturn in the slope of the superconducting \emph{H--T} phase boundary suggests the presence of a large supply of superconducting states accessible via magnetic field.  We stipulate that our observations suggest that the rapid reduction of the upper critical field with increasing temperature (concomitantly with the superfluid density) is a consequence of thermal fluctuations involving these states.  

In order to gain a deeper understanding into the nature of the superconducting state, we explored a scenario of Cooper pairs in the weakly-coupled limit with finite center-of-mass momentum.  By utilizing a theoretical approach developed by A. Gurevich \cite{Gurevich}, we find that even if the above model is not fully justifiable, it is still consistent with our data, since we observe the emergence of $Q \neq 0$ at $T \leq 2.5$ K.  We thus conclude that this observed exotic behavior in the superconducting phase diagram of FeTe:O$_{x}$ might be a consequence of the strongly-correlated nature of this particular system.
  
As our outlook into the future, we suggest that the recent discoveries of new kinds of iron chalcogenide superconductors, such as (Tl,Rb)$_{y}$Fe$_{2-x}$Se$_{2}$ which are marked by a significantly-enhanced electronic exchange couplings \cite{Kotliar,Liu} will offer a new testing ground for observing potentially exotic $H - T$ phase behaviors at high magnetic fields, while at the same time necessitating a better theoretical understanding of iron-chalcogenide electronic structure.     

\ack

The work at Brookhaven National Laboratory was supported by the Office of Science, U.S. Department of Energy, Materials Sciences and Engineering Division, under Contract No. DE-AC02-98CH10886.  A portion of this work was performed at the National High Magnetic Field Laboratory, which is supported by NSF Cooperative Agreement No. DMR-1157490 by the State of Florida, and by the DOE.  I.K.D. wishes to thank Vyacheslav Solovyov and Silvia Haindl for critical reading of the manuscript and helpful suggestions.


\section*{References}
\begin{thebibliography}{10}

\bibitem{Kamihara}
Kamihara Y, Watanabe T, Hirano M, and Hosono H, J. Am. Chem. Soc. \textbf{130}, 3296 (2008).

\bibitem{Mazin}
Mazin I I, Nature \textbf{464}, 183 (2010).

\bibitem{Zhao}
Zhao J, Huang Q, De La Cruz C, Li S, Lynn J W, Chen Y, Green M A, Chen G F, Li G, Li Z, Luo J L, Wang N L, and Dai P, Nature Mater. \textbf{7}, 953 (2008).

\bibitem{Lester}
Lester C, Chu J-H, Analytis J G, Capelli S C, Erickson A S, Condron C L, Toney M F, Fisher I R, and Hayden S M, Phys. Rev. B \textbf{79}, 144523 (2009).

\bibitem{Nandi}
Nandi S, Kim M G, Kreyssig A, Fernandes R M, Pratt D K, Thaler A, Ni N, Bud\'{}ko S L, Canfield P C, Schmalian J, McQueeney R J, and Goldman A I, Phys. Rev. Lett. \textbf{104}, 057006 (2010).

\bibitem{Chauviere}
Chauvi\`ere L, Gallais Y, Cazayous M, Measson M A, Sacuto A, Colson D, and Forget A, Phys. Rev. B \textbf{82}, 180521(R) (2010).

\bibitem{Hardy}
Hardy F, Burger P, Wolf T, Fisher R A, Schweiss P, Adelmann P, Heid R, Fromknecht R, Eder R, Ernst D, L\"{o}hneysen H v, and Meingast C, Eur. Phys. Lett. \textbf{91}, 47008 (2010).

\bibitem{Tinkham}
Tinkham M, \emph{Introduction to Superconductivity} (Second Edition, Dover Publications, Inc., Mineola, New York, 1996).

\bibitem{DeGennes}
De Gennes P G, \emph{Superconductivity of Metals and Alloys} (Westview Press, 1999).

\bibitem{Werthamer}
Werthamer N R, Helfand E, and Hohenberg P C, Phys. Rev. \textbf{147}, 295 (1966).

\bibitem{Clogston}
Clogston A M, Phys. Rev. Lett. \textbf{9}, 266 (1962); Chandrasekhar B S, Appl. Phys. Lett. \textbf{1}, 7 (1962).

\bibitem{Khim}
Khim S, Kim J W, Choi E S, Bang Y, Nohara M, Takagi H, and Kim K H, Phys. Rev. B \textbf{81}, 184511 (2010).

\bibitem{Cho}
Cho K, Kim H, Tanatar M A, Song Y J, Kwon Y S, Coniglio W A, Agosta C C, Gurevich A, and Prozorov R, Phys. Rev. B \textbf{83}, 060502(R) (2011).

\bibitem{Weidong}
Si W, Jie Q, Wu L, Zhou J, Gu G, Johnson P D, and Li Q, Phys. Rev. B \textbf{81}, 092506 (2010).

\bibitem{QL}
Li Q, Si W, and Dimitrov I K, Rep. Prog. Phys. \textbf{74}, 124510 (2011) and references therein.

\bibitem{Ivo1}
In order to avoid certain artifacts which may potentially arise in the data analysis from overestimation of ${\rho}_{0}(T)$ from the polynomial fit, we obtained the superconducting phase diagram of FeTe:O$_{x}$ also utilizing a much more rigorous criterion, which assumes that ${\rho}_{0}(T)$ remains constant below 7 K (figure 1(c)).  The resultant ${H}_{c2}^{\perp c}(T)$ curve is shown in the inset of figure 2 and we note that despite the fact that it is down-shifted with respect to the ${H}_{c2}^{\perp c}(T)$ obtained using the standard scheme, both phase diagrams show the same signature behavior.

\bibitem{Ivo2}
The normal state below $T \approx$  5 K could not be reached fully due to the very high ${H}_{c2}(T)$ of the sample.  The low-field data ($H \sim$ 0 -- 9 T) were confirmed by means of temperature scans performed in a Physical Property Measurement System by Quantum Design. 

\bibitem{Nakayama}
Nakayama K, Sato T, Richard P, Kawahara T, Sekiba Y, Qian T, Chen G F, Luo J L, Wang N L, Ding H, and Takahashi T, Phys. Rev. Lett. \textbf{105}, 197001 (2010).

\bibitem{Homes}
Homes C C, Akrap A, Wen J S, Xu Z J, Lin Z W, Li Q, and Gu G D, Phys. Rev. B \textbf{81}, 180508(R) (2010).

\bibitem{Luetkens}
Luetkens H, Klauss H-H, Khasanov R, Amato A, Klingeler R, Hellmann I, Leps N, Kondrat A, Hess C,
K\"{o}hler A, Behr G, Werner J, and B\"{u}chner B, Phys. Rev. Lett. \textbf{101}, 097009 (2008).

\bibitem{Ku}
Lee C-C, Yin W-G, and Ku W, Phys. Rev. Lett. \textbf{103}, 267001 (2009).

\bibitem{Kotliar}
Yin Z P, Haule K, and Kotliar G, Nat. Mater. \textbf{10}, 932 (2011) and references therein.

\bibitem{Kato}
Kato T, Mizuguchi Y, Nakamura H, Machida T, Sakata H, and Takano Y, Phys. Rev. B \textbf{80}, 180507 (2009).

\bibitem{Hanaguri}
Hanaguri T, Niitaka S, Kuroki K, Takagi H, Science \textbf{328}, 474 (2010).

\bibitem{Larkin}
Larkin A I, and Ovchinnikov Yu N, Zh. Eksp. Teor. Fiz. \textbf{47} 1136 (1964).

\bibitem{Fulde}
Fulde P, and Ferrell R A, Phys. Rev. \textbf{135}, A550 (1964).

\bibitem{Matsuda}
Matsuda Y, and Shimahara H, J. Phys. Soc. Jpn \textbf{76}, 051005 (2007) and references therein.

\bibitem{Tanatar}
Tanatar M A, Ishiguro T, Tanaka H, and Kobayashi H, Phys. Rev. B \textbf{66}, 134503 (2002); Uji S, Shinagawa H, Terashima T, Yakabe T, Terai Y, Tokumoto M, Kobayashi A, Tanaka H, and Kobayashi H, Nature \textbf{410}, 908 (2001); Balicas L, Brooks J S, Storr K, Uji S, Tokumoto M, Tanaka H, Kobayashi H, Kobayashi A, Barzykin V, and Gor'kov L P, Phys. Rev. Lett. \textbf{87}, 067002 (2001); Houzet M, Buzdin A, Bulaevskii L, and Maley M, Phys. Rev. Lett. \textbf{88}, 227001 (2002); Uji S, Terashima T, Nishimura M, Takahide Y, Konoike T, Enomoto K, Cui H, Kobayashi H, Kobayashi A, Tanaka H, Tokumoto M, Choi E S, Tokumoto T, Graf D, and Brooks J S, Phys. Rev. Lett. \textbf{97}, 157001 (2006); Izawa K, Yamaguchi H, Sasaki T, and Matsuda Y, Phys. Rev. Lett. \textbf{88}, 027002 (2002). 

\bibitem{Gloos}
Gloos K, Modler R, Schimanski H, Bredl C D, Geibel C, Steglich F, Buzdin A I, Sato N, and Komatsubara T, Phys. Rev. Lett. \textbf{70}, 501 (1993); Yin G, and Maki K, Phys. Rev. B \textbf{48}, 650 (1993); Tachiki M, Takahashi S, Gegenwart P, Weiden M, Lang M, Geibel C, Steglich F, Modler R, Paulsen C, Onuki Y, Z. Phys. B \textbf{100}, 369 (1996); Buzdin A I, and Kachkachi, Phys. Lett. A \textbf{225}, 341 (1997); Mitrovi\'c V F, Horvati\'c M, Berthier C, Knebel G, Lapertot G, and Flouquet J, Phys. Rev. Lett. \textbf{97}, 117002 (2006); Koutroulakis G, Stewart, Jr. M D, Mitrovi\'c V F, Horvati\'c M, Berthier C, Lapertot G, and Flouquet J, Phys. Rev. Lett. \textbf{104}, 087001 (2010). 

\bibitem{Gurevich}
Gurevich A, Phys. Rev. B \textbf{82}, 184504 (2010).

\bibitem{Xia}
Xia Y, Qian D, Wray L, Hsieh D, Chen G F, Luo J L, Wang N L, and Hasan M Z, Phys. Rev. Lett. \textbf{103}, 037002 (2009).

\bibitem{Kogan}
Kogan V G, Phys. Rev. Lett. \textbf{89}, 237005 (2002).

\bibitem{Kidszun}
Kidszun M, Haindl S, Thersleff T, H\"{a}nisch J, Kauffmann A, Iida K, Freudenberger J, Schultz L, and Holzapfel B, Phys. Rev. Lett. \textbf{106}, 137001 (2011).

\bibitem{Blatter}
Blatter G, Geshkenbein V B, and Larkin A I, Phys. Rev. Lett. \textbf{68}, 875 (1992); Iida K, H\"{a}nisch J, Thersleff T, Kurth F, Kidszun M, Haindl S, H\"{u}hne, Schultz L, and Holzapfel B, Phys. Rev. B \textbf{81}, 100507(R) (2010).

\bibitem{Liu}
Liu Z-H, Richard P, Xu N, Xu G, Li Y, Fang X C, Jia L L, Chen G F, Wang D M, He J B, Qian T, Hu J P, Ding H, and Wang S C, Phys. Rev. Lett. \textbf{109}, 037003 (2012).


\end{thebibliography}
\end{document}